# An Information Analysis on Modeling Interaction Effects in Logistic Regression

Jiun-Wei Liou, Michelle Liou, Philip E. Cheng

Institute of Statistical Science, Academia Sinica, Taipei, Taiwan

and

Chin-Chiuan Lin

Department of Business Administration

Kung-Shan University, Tainan, Taiwan

**Abstract**

The Akaike information criterion (AIC) is commonly used to select a logistic regression model for optimal prediction of a binary response by a specified family of models. It however lacks a convincing method of prescribing a proper family of models using the desired predictors and their interaction effects. For an alternative approach to model selection, we propose a direct selection scheme which first identifies the indispensable regressors as main-effect predictors, then examines significant interaction effects between the selected predictors such that a logistic model is constructed. The two-step selection scheme is formulated by testing for valid information identity between the response and the predictors, from which the most parsimonious logistic model is derived from the least set of indispensable predictors and interaction effects. As a byproduct, the minimum AIC model is easily found in a neighborhood of the selected model. The scheme is employed to yield the logistic model for predicting the acquisition of professional licenses in a survey of employed youth workers.

**Correspondence**: Michelle Liou, Institute of Statistical Science, Academia Sinica, Taipei, Taiwan. Email: mliou@stat.sinica.edu.tw

# 1. Introduction

Analysis of contingency tables is commonly processed using log-linear and logistic models. Selection and interpretation of these models have been widely discussed in the literature (Agresti, 2013; Bishop et al., 2007; McCullagh and Nelder, 1989). In social and health sciences, researchers have frequently elaborated on the main effects exclusively in the logistic regression analysis especially when the number of regressors is moderately large. For example, the backward stepwise selection procedure was employed along with the minimum Akaike information criterion (AIC) when selecting social and ecological factors to address the declination problem of African lion populations (Hazzah et al., 2009). The prevalence of emotional problems of children was studied for those raised by same-sex parents as compared with those by opposite-sex parents by controlling for potential confounders such as parents' psychological distress and children's peer stigmatization (Sullins, 2015). Further, the associations between the Bisphenol A (an organic compound used in the synthesis of plastics and epoxy resins) exposure and cardiovascular disease (or diabetes) were reported by adjusting for health conditions and major demographic factors (Casey and Neidell, 2013). In these studies, research interest was mainly centered on the unique contribution of each individual regressor without considering possible contribution of interaction effects when specifying the model for scientific interpretation.

The interaction between two regressors implies "moderation" in that the effect of one regressor on the response variable is differentially moderated by the other. That is, the moderation effect could increase or decrease the association between the regressors and the response variable and confound data interpretation. It is indeed difficult to interpret the main effect of a regressor when a significant two- or multi-way interaction effect is also present in the model, rendering a case comparable to the Simpson paradox (Simpson, 1951). While models consisting of purely main effects are popular or preferable in the literature, the use of interaction effects in logistic regression models was also remarked (Jaccard, 2001), and inevitably related to the well-known issue of model selection. In principle, the AIC is designed to minimize the loss of information by penalizing the inclusion of parameters that contribute little to the model likelihood (Akaike, 1974; Burnham and Anderson, 2004; Linhart and Zucchini, 1986; Sugiura, 1978). Given a set of regressors, it follows that the minimum AIC offers parsimonious models governed by Occam's razor for data interpretation and prediction. The search for the minimum AIC model is commonly initiated using a family of competing models, but executed without guidelines for incorporating potential interactions among regressors in the model.

Instead of relying on an optimization criterion, this study will construct a logistic regression model through estimating the mutual information (MI) of variables based on the multivariate multinomial distribution of categorical variables. An information identity presents a geometric account of mutual information in terms of orthogonal components of lower- to higher-order interaction effects (Cheng et al., 2007). We will formulate a two-step selection scheme that begins with the removal of dispensable regressors by evaluating the MI between the target and each regressor conditional on the remaining regressors. In the second step, the scheme selects an MI identity to yield the least significant higher-order interactions by testing between the target and the selected regressors from the first step. The proposed two-step scheme therefore constructs the most parsimonious logistic model using the indispensable regressors (main effects) and the least interaction effects among models selected by any existing criteria. The scheme will be illustrated using a real data set collected from the Youth Labor Employment

Survey administered by the Center of Survey Research Data Archive (SRDA), Academia Sinica in 2008. The survey conducted 183 questions focused on a wide range of work characteristics of a random sample of 4,012 employed youth workers. Among the 183 survey questions, responses on six key questions about individual's background information were selected for their potential relevance to the status of acquiring professional licenses, that is, gender, education, school major, employment counseling, workplace training, and advanced study (Bynner and Parsons, 2002; Dekker et al., 2002; Kleiner and Krueger, 2010; McCullagh and Nelder, 1989).

This study is layed out as follows. In Section 2, a review is devoted to the conventional logistic regression analysis on categorical variables, and to the MI identity based on the discrete multinomial distribution and its relationship with log-linear models. A system of two-step selection scheme is proposed for identifying important regressors based on conditional mutual information (CMI), followed by constructing an MI identity with least number of interaction effects. In this aspect, the best MI identity implies the selection of the most parsimonious logistic model. In Section 3, the survey data set is described, followed by a systematic analysis using the conventional logistic regression method and the AIC. The difficulty in selecting models including interaction effects is discussed. In Section 4, the proposed two-step scheme is applied to the survey data set and the most parsimonious model is decided against the best AIC model. A simulation study is performed to help illustrating the proposed two-step procedure. In summary, the proposed scheme identifies the main-effect regressors and the interaction effects by testing an MI identity such that a concise logistic model is constructed. It generally yields the most parsimonious model, and a few neighbor models can be checked to identify the minimum AIC model with ease. As a final remark, it is expected to extend the proposed information approach to selecting general logistic models using both categorical and continuous regressors.

## 2. Logistic Model and Information Identity

The classical analysis of partitioning chi-squares and testing for conditional independence in a 3-way table inspired the development of the log-linear model (Goodman, 1970; Kullback, 1968), the generalized linear models (Nelder and Baker, 1972), and analysis of binary outcomes using categorical regressors (Bishop et al., 2007; Goodman and Kruskal, 1979). A basic link between the analysis of log-linear model and mutual information was recently discussed with testing conditional independence in a 3-way contingency table (Cheng et al., 2006). In this section, a basic connection between the MI identity and the logistic models or log-linear models is elaborated using 3- and 4-way contingency tables.

### 2.1 Basic Logistic Models

Let (X, Y, Z) denote a three-way $I \times J \times K$ contingency table with the joint probability density function (pdf) $f_{X,Y,Z}(i,j,k), i = 0, \cdots I; j = 0, \cdots J; k = 0, \cdots K;$ which denotes the expected frequency in the $(i,j,k)$ cell under the assumption of (Poisson-) multinomial distribution (Lehmann et al., 1986). Let Y be the binary response variable of interest, in short, the target, with $J = 1$ for simplicity. To predict the target Y based on X and Z, a logistic model with purely main effects can be expressed as

$$\text{logit}\big[f_{Y|X,Z}(Y=1 \mid X=i, Z=k)\big] \equiv \log\left[\frac{f_{Y|X,Z}(Y=1|i,k)}{f_{Y|X,Z}(Y=0|i,k)}\right] = \beta_0 + \beta_i^X + \beta_k^Z, \quad (1)$$

where $f_{Y|X,Z}(Y = 1|i,k)/f_{Y|X,Z}(Y = 0|i,k)$ denotes the odds of Y = 1 versus Y = 0 when the regressors are evaluated at (X = $i$, Z = $k$). Model (1) is equivalent to the log-linear model {XY, YZ, XZ} without interaction among the three variables (Agresti, 2013; McCullagh and Nelder, 1989). With three regressors X, Z, and W, the logistic model of the main effects plus the XZ interaction effect can be expressed as

$$\text{logit}[f_{Y|X,Z,W}(Y = 1 \mid X = i, Z = k, W = l)] \equiv \log\left[\frac{f_{Y|X,Z,W}(Y=1|i,k,l)}{f_{Y|X,Z,W}(Y=0|i,k,l)}\right]$$
$$= \beta_0 + \beta_i^X + \beta_k^Z + \beta_{ik}^{XZ} + \beta_l^W. \quad (2)$$

The equivalent log-linear model to (2) is {XYZ, WY, XWZ}, which extends model (1) to the case of no interaction between Y, W and XZ. In the literature, there are two basic tools for examining a logistic model, as has been supported in many statistics software. First, the inclusion or exclusion of a regressor is commonly inspected using the Type III likelihood ratio (LR) statistic which tests the unique contribution of a regressor when all other regressors are already in the model. Second, given the selected regressors, the logistic model that maximizes the prediction accuracy is provided with the minimum AIC, which yields the greatest log-likelihood by using the least number of parameters. As an alternative to the standard approach, we will introduce a geometric analysis of selecting indispensable regressors and constructing the most parsimonious logistic model by testing a proper MI identity with the selected regressors.

**2.2 Geometric Information Analysis**

Given a contingency table, the association between the marginal variables can be described using the MI. The MI of a few variables can be expressed as sum of MI and CMI terms, which are components, each associated with a subset of the variables, in orthogonal decomposition (Cheng et al., 2007; Kullback, 1968). In a 2 x 2 table, the MI is used to characterize the invariant Pythagorean law of relative entropy, which is the geometric analog of the Pearson chi-square test for two-way independence (Cheng et al., 2008).

**2.2.1 Information Identity**

As mentioned, an information identity of variables in a contingency table is used to characterize the geometry of association between categorical variables by orthogonal decomposition. This geometry can be introduced with the Shannon entropy which defines the basic information identity of three variables as

$$H(X) + H(Y) + H(Z) = I(X; Y; Z) + H(X; Y; Z),$$

where

$$H(X; Y; Z) = -\sum_{i,j,k} f_{X,Y,Z}(i,j,k) \log[f_{X,Y,Z}(i,j,k)]$$

is the joint entropy of (X, Y, Z), and

$$I(X; Y; Z) = \sum_{i,j,k} f_{X,Y,Z}(i,j,k) \log\left(\frac{f_{X,Y,Z}(i,j,k)}{f_X(i)f_Y(j)f_Z(k)}\right) \qquad (3)$$

is the MI of (X, Y, Z) with the marginal pdfs $f_X$, $f_Y$ and $f_Z$ (Cover and Thomas, 2006; Kullback and Leibler, 1951). The MI, being the KL-divergence, defines an analog of the "projection" from the observed data joint pdf to the product space of marginal pdfs, that is, the parameter space of the null hypothesis of independence (Cheng et al., 2010). The MI of three variables is usually expressed as a sum of three quantities, say,

$$I(X; Y; Z) = I(X; Z) + I(Y; Z) + I(X; Y|Z). \tag{4}$$

The last summand in (4) is defined as the conditional mutual information (CMI)

$$I(X; Y|Z) = \sum_k \left\{ \sum_{ij} f_{XY|Z}(i,j|k) \log \left[ \frac{f_{XY|Z}(i,j|k)}{f_{X|Z}(i|k) f_{Y|Z}(j|k)} \right] \right\}.$$

Equation (4) is a basic information identity in three variables, where Z is termed a conditioning variable, and an exchange of any two variables would yield an equivalent identity. Using the multinomial sampling distribution, the sample analogs of the MI and CMI terms in (4) are defined by replacing the pdfs with their sample frequency estimates, which are the maximum likelihood estimates (MLEs) under the hypotheses of independence (Lehmann et al., 1986). The factor $2N$ (with sample size $N$) is attached to the sample MI and CMI estimates for valid approximations to the chi-square distributions, defined as

$$\hat{I}(X; Y; Z) = 2N \sum_{i,j,k} \hat{f}_{X,Y,Z}(i,j,k) \log \left( \frac{\hat{f}_{X,Y,Z}(i,j,k)}{\hat{f}_X(i) \hat{f}_Y(j) \hat{f}_Z(k)} \right),$$

where $\hat{f}_{X,Y,Z}(i,j,k)$ denotes the sample joint density estimate in the $(i,j,k)$-th cell, and other notations are defined by analogy. Thus, the sample analogs of the terms in (4) are asymptotically chi-square distributed with IJK − I − J − K + 2 degrees of freedom (*df*s) on the left-hand side, and (I - 1)(K − 1), (J − 1)(K − 1), and (I − 1)(J − 1)K *df*s on the right-hand side, respectively. In particular, the sample MI $\hat{I}(X; Y; Z)$ measures the same deviance as does the main-effect log-linear model {X, Y, Z}. It is essential that the CMI I(X; Y|Z) in (4) can be further decomposed as the sum of two orthogonal terms:

$$I(X; Y|Z) = Int(X; Y; Z) + Par(X; Y|Z). \qquad (5)$$

By (5), the sample analog $\hat{I}(X; Y|Z)$ yields the same deviance when testing for the log-linear model {XZ, YZ}, and likewise, the sample analog $\widehat{Int}(X; Y; Z)$ measures the same deviance as the log-linear model {XY, YZ, XZ} (Bishop et al., 2007; Deming and Stephan, 1940). The sample analog $\widehat{Par}(X; Y|Z)$, obtained as the difference $\hat{I}(X; Y|Z)$ - $\widehat{Int}(X; Y; Z)$, estimates the uniform association between X and Y across the levels of Z. Traditionally, the term $Par(X; Y|Z)$ is coined the partial association between X and Y given Z. It is worth noting that the additional contribution of the term {XY}, toward reducing the LR deviance of the log-linear model {XZ, YZ}, is accounted for the three-way partial association $\widehat{Par}(X; Y|Z)$, instead of the two-way effect $\hat{I}(X; Y)$.

The sample estimates $\hat{I}(X; Y|Z)$, $\widehat{Int}(X; Y; Z)$ and $\widehat{Par}(X; Y|Z)$ of equation (5) are asymptotically chi-square distributed with $(I - 1)(J - 1)K$, $(I - 1)(J - 1)(K - 1)$ and $(I - 1)(J - 1)$ *df*s, respectively. When the test for the log-linear model {XZ, YZ} is significant, it is crucial that the two-step LR tests are applicable to the summands in (5). That is, the usual test level $\alpha$, say, 0.05, should be divided into a pair of two levels ($\alpha_1$, $\alpha_2$), such that $\widehat{Int}(X; Y; Z)$ and $\widehat{Par}(X; Y|Z)$ are tested using $\alpha_1$ and $\alpha_2$, respectively, where $\alpha_1 + (1 - \alpha_1)\alpha_2 = \alpha$ (Cheng et al., 2010). Meanwhile, it follows from (4) and (5) that a useful information identity for illustrating Y in terms of X and Z is

$$I(\{X, Z\}; Y) = I(X; Y; Z) - I(X; Z)$$
$$= I(Y; Z) + Int(X; Y; Z) + Par(X; Y|Z). \qquad (6)$$

By definition, the sample analog $\hat{I}(\{X, Z\}; Y)$ gives the LR deviance of the log-linear model {Y, XZ}, and the terms of equation (6) present an MI identity for illustrating Y by the variables X and Z. Extensions of (6) to the case of four or more variables are straightforward (Cheng et al., 2007), for example,

$$\begin{aligned} I(\{W, X, Z\}; Y) &= I(Y; Z) + I(W; Y|Z) + I(X; Y|\{W, Z\}) \\ &= I(Y; Z) + Int(W; Y; Z) + Par(W; Y|Z) \\ &\quad + Int(X; Y; \{W, Z\}) + Par(X; Y|\{W, Z\}). \end{aligned} \quad (7)$$

The sample analog $\hat{I}(\{W, X, Z\}; Y)$ of equation (7) measures the deviance of the log-linear model {Y, WXZ}, and it illustrates Y by W, X, and Z, through using a decomposed MI identity, where each CMI term is further decomposed as a pair of interaction and partial association terms. Note that equivalent MI identities to (7) can be defined by exchanging X and Z (or, X and W). Inference for Y in a contingency table can be examined using MI identities as analogies of (6) and (7), and the two-step LR tests discussed in (5) and (6) are applicable to the decomposed terms in (7). These basic analyses will be useful for the construction of logistic models.

### 2.2.2 A Selection Scheme

A primary goal of this study is to develop a selection scheme for a logistic model which illustrates a binary target Y by its covariates in a finite-dimensional contingency table. The selection scheme consists of two parts. The first part removes dispensable regressors for the target Y using a sequence of CMI estimates, and the second part selects a proper information identity by testing for significant interaction effects between Y and the selected regressors $X_{(t)}$ for $t = 1, \ldots, k^*$.

Suppose that there are k covariates associated with the binary target Y. A few notations are defined in order. Let $R^{(0)}$ denote the set of all regressors, and let $R^{(k)}$ denote the empty set. For $t = 1, \cdots, k$, define $\hat{I}(Y; X_{(t)}| R^{(t)})$ to be the $t$-th highest-order CMI between the target Y and the regressor $X_{(t)}$, conditional on $R^{(t)} = R^{(t-1)} \setminus X_{(t)}$, the regressor set at the $t$-th selection stage. The first part of the selection scheme is accomplished when, say, $k - k_0 + 1$ (= k*) regressors remain in effect (or selected) after deleting $k_0 - 1$ insignificant regressors based on testing the sample CMI statistics. The second part of the scheme is designed to rearrange the selected regressors according to a finite sequence of the decomposed MI and CMI terms (in a proper information identity) between Y and the selected regressors, which will lead to valid logistic models. This begins with selecting the least significant interaction term between Y and a regressor (with the largest *p*-value in the two-step LR test for equation (5)) conditional on other available regressors. Continue the selection with subsequent decompositions into pairs of interaction and partial association terms for the remaining regressors by the same rule. The scheme of selecting the set of useful regressors and a proper information identity of association effects is displayed in Figure 1 below.

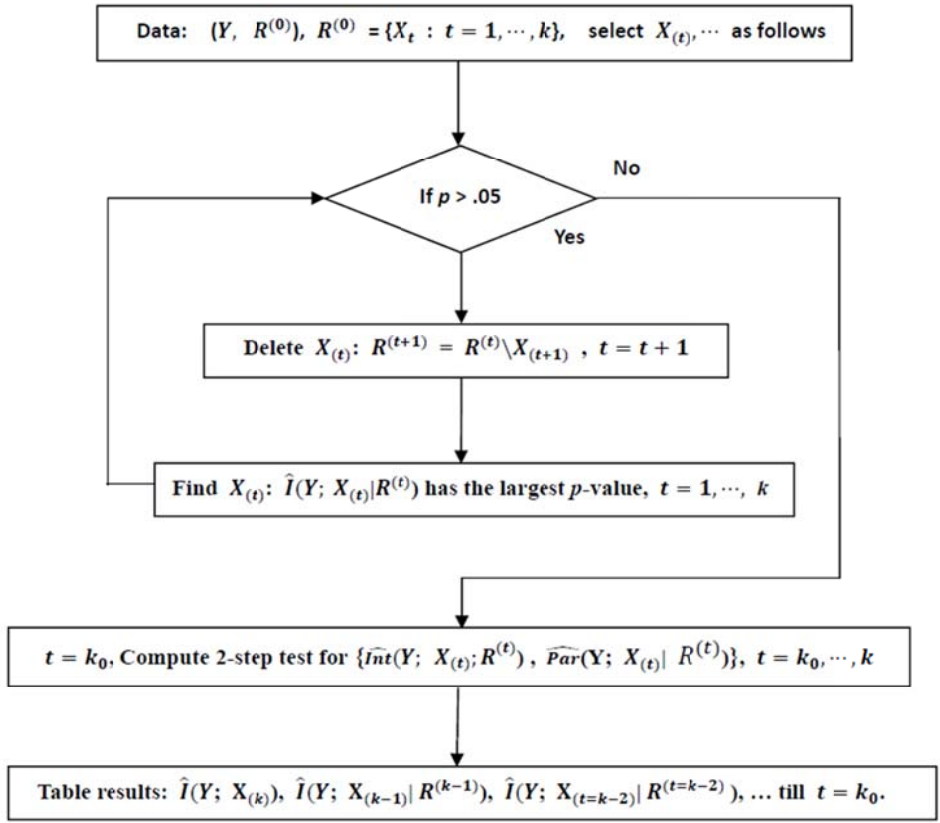

Figure 1: Variable and model selection scheme by mutual information.

The scheme yields a table of decomposed MI and CMI terms of the selected information identity. Details will be exemplified using the SRDA dataset, for which the decomposed information identity is presented in Table 1 of Section 4.

### 3. Logistic Regression in Practice

A brief review of the standard procedure of testing for valid logistic models is discussed together with the AIC models using the survey data set. In many job markets, the status of holding one or more professional licenses has been commonly regarded as beneficial for entering prestigious levels of employment in the labor market as well as for the entry-level wages and continuing employment status. In the early 2000s, there were more than fifty types of professional licenses certified by the Skill Evaluation Center administered under the Workforce Development Agency of Taiwan Ministry of Labor, along with a few private associations and testing centers. It is of interest to evaluate the status of acquiring professional licenses as a function of socio-demographic factors among the employed youth. In the SRDA data set, 2,347 out of 4,012 employed youth workers have acquired at least one professional license. A binary variable is used to define the status of having at least one professional license or not, denoted by L = 1 or 0. Six binary regressors, each coded as "1 or 0", are found useful for explaining the license status, including "A = 1, having *advanced* study on employment, and 0 otherwise", "T = 1, having attended at least a job *training* program, and 0 otherwise", "C = 1, having attended an employment *counseling* program, and 0 otherwise", "E = 1, having a bachelor's degree or higher *education*, and 0 otherwise", "M = 1, *majoring* in engineering related fields, and 0 otherwise" and "G = 1 for male in *gender*, and 0 for female". Thus, the data consists of a contingency table with seven binary variables (Note: The data set is available from the corresponding author).

In this contingency table, a useful class of descriptive statistics is the set of logarithmic odds ratio (OR) estimates between the target and individual binary explanatory variables in the six $2 \times 2$ raw data marginal tables. These logarithmic OR estimates are of interest because they closely resemble the estimated regression coefficients of the corresponding regressors in the main-effect logistic model. For example, the logarithmic OR estimates of the two $2 \times 2$ marginal tables between the target and the binary regressors "C" and "E" are 0.046 ($p$ = .508) and -0.043 ($p$ = .633), respectively. These two sample OR estimates are insignificant when testing for the null hypothesis of independence between the target and each individual regressor. It turns out that the acquired main-effect logistic model is

$$\text{Logit}[f(L|A, C, E, G, M, T)] = 0.688 - 0.350A + 0.025*C - 0.325G + 0.479T$$
$$- 0.252M - 0.088*E. \qquad (8)$$

The regressors with insignificant parameter estimates (marked with asterisks) in (8) are shown as C and E, which are also presented with insignificant Type III LR test statistics 0.128 ($p$ = .720) and 0.793 ($p$ = .373), respectively. Meanwhile, the LR residual deviance for testing the goodness-of-fit of (8) is 81.661 ($df$ = 57, $p$ = 0.018). The lack-of-fit of (8) suggests that inclusion of interaction effects between regressors may be necessary before justifying the exclusion of the regressor "C", or "E", or both, for the inference about the target L (McCullagh and Nelder, 1989).

For decades, the AIC has been a standard tool for the selection of generalized linear models, including the logistic regression model. The minimum AIC model is often preferred to other criteria due to its strength of achieving optimal prediction accuracy (Burnham and Anderson, 2004; Fahrmeir and Tutz, 2013; Kateri, 2014; Tutz, 2011). To search for the minimum AIC logistic model for the SRDA data set, inclusion of interaction effects to (8) is considered. Backward- and forward-stepwise selections of generalized linear models have been commonly employed for analyzing logistic regression (Towner and Luttbeg, 2007). The target L and six regressors can be input into the logistic regression analysis in SAS or SPSS by including the main and all possible $2^{nd}$ and higher order interactions. Alternatively, a finite sequence of stepwise selections of models using different levels of interaction effects can be examined in order and the minimum AIC model is found to be

$$\text{Logit}[f(L|A, C, E, G, M, T)] = 0.814 - 0.354A + 0.361T - 0.381G - 0.439M - 0.715E$$
$$+ 0.028*C + 0.589EG + 0.838EM + 0.261*MT. \quad (9)$$

Model (9) yields the log-likelihood -139.668, the minimum AIC estimate 299.336 and LR residual deviance 52.552 ($df = 54$, $p = 0.530$). It includes two insignificant parameter estimates for the main effect C ($p = 0.689$) and the interaction effect MT ($p = 0.056$); and, it also presents four insignificant Type III LR chi-square statistics on the main effects C ($p = 0.689$), G ($p = 0.397$), M ($p = 0.309$) and E ($p = 0.998$). These test statistics might not provide valid inference on the removal of any regressor, except that the factor C is consistently insignificant. If the insignificant MT effect is removed from (9), a more concise model is obtained as

$$\text{Logit}[f(L|A, C, E, G, M, T)] = 0.761 - 0.354A + 0.484T - 0.731E - 0.384G - 0.340M$$
$$+ 0.023*C + 0.587EG + 0.870EM. \quad (10)$$

Model (10) yields the log-likelihood -141.502, AIC estimate 301.003, residual deviance 56.219 ($df = 55$, $p = 0.429$), and similar Type III chi-square statistics to those of (9), among which the main effect C is also insignificant ($p = 0.742$).

Results in (9) and (10) strongly suggest that the variable C may be removed. Indeed, it leads to a more parsimonious model by deleting C, that is,

$$\text{Logit}[f(L|A, E, G, M, T)] = 0.769 - 0.354A - 0.731E - 0.384G - 0.340M$$
$$+ 0.485T + 0.587EG + 0.870EM. \quad (11)$$

Model (11) yields all significant parameter estimates with the log-likelihood -84.029, the AIC estimate 184.058 and the residual deviance 29.060 ($df = 24$, $p = 0.218$). Meanwhile, the minimum AIC model using the five regressors can be easily found from a few neighboring models of (11), that is,

$$\text{Logit}[f(L|A, E, G, M, T)] = 0.823 - 0.354A - 0.715E - 0.381G - 0.438M + 0.362T$$
$$+ 0.589EG + 0.839EM + 0.259*MT. \qquad (12)$$

Model (12) yields the minimum AIC estimate 182.444, the residual deviance 25.455 ($df = 23$, $p = 0.327$), and all significant parameter estimates except that the MT interaction estimate has a barely significant chi-square test statistic 3.608 ($p = 0.057$). The MT interaction term may be deleted from (12) such that model (11) is obtained, because the delta AIC between the models is less than 2.0 [32]. Thus, the above has presented a basic analysis of logistic regression which combines the AIC and type III analysis for the SRDA data set. In the next section, an information theoretical approach to modeling logistic regression using the selection scheme of the previous section will be examined with the same dataset.

## 4. Information Geometry in Logistic Regression

For notational ease, let Y and the set {W, X, Z} in (7) be replaced by the target L and the set of regressors {A, C, E, G, M, T}, respectively. We will employ the first part of the selection scheme in Section 2.2.2 to select the desired predictors for L. Among six similar $6^{th}$ order CMI terms, the least significant term is $\hat{I}(L; C|\{T, G, M, E, A\}) =$ 27.263 ($df = 32, p = 0.705$), and the next is $\hat{I}(L; E|\{T, G, M, C, A\}) = 55.296$ ($df = 32, p = 0.006$), which is significant. The variable C is found as the first redundant regressor for the target L. By analogy, the next least significant CMI estimate among the remaining five regressors {A, T, G, M, E} identifies the variable A with $\hat{I}(L; A|\{T, G, M, E\}) = 35.727$ ($df = 16, p = 0.003$), which is however significant. This proves that C is the only redundant regressor, and the other five regressors {A, T, G, M, E} are retained for the target L, giving a more confirmative selection of regressors than the previous Type III analysis in (8) to (10). Indeed, deleting the variable C manifests the background fact that the survey was conducted among the employed youth workers, so that the employment counseling effect designed (mainly for the unemployed youths) for the acquirement of professional licenses is not evident in the SRDA data.

Using the selected regressors {A, T, G, M, E}, the second part of the selection scheme in Figure 1 is implemented to rearrange the interaction terms $\widehat{Int}(L, X_{(t)}, R^{(t)})$ according to their $p$-values. It is found that $\widehat{Int}(L; A; \{T, G, M, E\})$ (= 17.158) yields the largest $p$ value ($df = 15, p = 0.310$) among five similar interaction estimates. Putting aside the regressor A, the selection scheme continues to yield the next interaction estimate $\widehat{Int}(L; T; \{G, M, E\})$ (= 11.258) which has the largest $p$ value ($df = 7, p = 0.128$) among four alike estimates. The scheme continues with the variables G and M, and stops with the two-way MI term, i.e., $\hat{I}(L; E)$. For the present data, the resulting sequence of the CMI estimates, the decomposed interaction estimates and partial association estimates are presented in the order A → T→ G→ M→ E, which are listed in Table 1. As illustrated, the first line in Table 1 gives the pair of decomposed estimates, where the interaction estimate is the least significant with the largest $p$ value 0.310.

Table 1: Sequential decomposed CMI terms of the MI identity in (13) (α = 0.05)

| MI, CMI Terms | $\hat{I}(L; X_{(t)}\mid R^{(t)})$ | | | $\widehat{Int}(L; X_{(t)}; R^{(t)})$ | | | $\widehat{Par}(L; X_{(t)}\mid R^{(t)})$ | | |
|---|---|---|---|---|---|---|---|---|---|
| | CMI estimate | df | p-value | Interaction | df | p-value | Partial Assoc. | df | p-value |
| $\hat{I}(L; A\mid T, G, M, E)$ | 35.727 | 16 | 0.003 | 17.158 | 15 | 0.310 ($\alpha_1 = 0.034$) | 18.569 | 1 | < 0.001 ($\alpha_2 = 0.017$) |
| $\hat{I}(L; T\mid G, M, E)$ | 71.421 | 8 | < 0.001 | 11.258 | 7 | 0.128 ($\alpha_1 = 0.031$) | 60.163 | 1 | < 0.001 ($\alpha_2 = 0.020$) |
| $\hat{I}(L; G\mid M, E)$ | 29.071 | 4 | < 0.001 | 7.854 | 3 | 0.049 ($\alpha_1 = 0.028$) | 21.217 | 1 | < 0.001 ($\alpha_2 = 0.022$) |
| $\hat{I}(L; M\mid E)$ | 52.907 | 2 | < 0.001 | 19.897 | 1 | < 0.001 ($\alpha_1 = 0.025$) | 33.010 | 1 | < 0.001 ($\alpha_2 = 0.025$) |
| $\hat{I}(L; E)$ | 0.228 | 1 | 0.633 | | | | | | |

Table 1 specifically lays out the orthogonal decomposition of the MI, between the target L and the five regressors, according to the selection scheme of Figure 1. This yields the desired MI which eliminates the insignificant higher-order interaction estimates. That is,

$$\hat{I}(\{A, T, G, M, E\}; L) = \widehat{Int}(L; A; \{T, G, M, E\})^* + \widehat{Par}(L; A\mid\{T, G, M, E\})$$
$$+ \widehat{Int}(L; T; \{G, M, E\})^* + \widehat{Par}(L; T\mid\{G, M, E\})$$
$$+ \widehat{Int}(L; G; \{M, E\})^* + \widehat{Par}(L; G\mid\{M, E\})$$
$$+ \widehat{Int}(L; M; E) + \widehat{Par}(L; M\mid E) + \hat{I}(L; E). \qquad (13)$$

According to the two-step LR test of the Pythagorean law illustrated with equation (6) of Section 2, three higher order interaction terms (shaded) in Table 1 are insignificant, respectively, as marked with asterisks in the MI identity (13). Moreover, these three insignificant interaction terms are altogether insignificant, i.e., $\widehat{Int}$(L; A; {T, G, M, E}) + $\widehat{Int}$(L; T; {G, M, E}) + $\widehat{Int}$(L; G; (M, E)) = 17.158 + 11.258 + 7.854 = 36.270 ($df$ = 25, $p$ = 0.068). This fact is unaffected by exchanging the order between {A, T, G} in Table 1 by the invariant information of the CMI between L and {A, T, G} conditional on {E, M}, yet the present decomposition is simple by the selection scheme of Figure 1. Thus, by deleting these three insignificant interaction terms in (13), the remaining main and interaction effects are exactly constructed to yield the desired logistic model consisting of five main effects and a unique ME interaction, which is fitted to a new six-way contingency table. As usual, a log-likelihood equation (e.g. SPSS) is solved to yield the following logit model with parameter estimates as

$$\text{Logit}[f(L|A, E, G, M, T)] = 0.745 - 0.330E - 0.356M + 0.889EM - 0.307G$$

$$+ 0.482T - 0.349A, \quad (14)$$

which is valid with residual deviance 37.198 ($df$ = 25, $p$ = 0.055) by the information analysis in Table 1. The parameter estimates in (14) are all significant, the significant EM parameter estimate is clearly supported by the significant interaction $\widehat{Int}$(L; M; E) (= 19.897, $df$ = 1; $p$ < 0.001). However, the significant estimate of E, like that in (11), cannot reflect the insignificant MI $\hat{I}$(L; E) (= 0.228, $df$ = 1; $p$ = 0.633) or the partial association estimate (by interchanging E and M) $\widehat{Par}$(L; E|M) (= 3.045, $df$ = 1; $p$ = 0.) in Table 1, which is comparable to the insignificant Type III effect in models (8) and (9). Indeed, the significant EM interaction estimate 0.889 is rather close to the logarithmic odds ratio estimate 0.939 of the raw data, i.e., the logarithmic ratio of "the odds ratio 1.926 of the 2 × 2 EM table at L =1" to "the counterpart odds ratio 0.753 at L = 0", in the three-way (L, M, E) raw data table.

The MI identity (13) can be used to construct other valid models. For example, if the interaction terms {EM, EGM} in Table 1 are retained by deleting only the two highest-order interaction terms. Then, a logit model is acquired with residual deviance $\widehat{Int}(L; A; \{T, G, M, E\}) + \widehat{Int}(L; T; \{G, M, E\}) = 17.158 + 11.258 = 28.416$ ($df = 22$, $p = 0.162$), that is,

$$\text{Logit}[f(L|A, E, G, M, T)] = 0.787 - 0.708E - 0.383M - 0.462G + 0.488T - 0.351A$$
$$+ 0.710*EM + 0.127*GM + 0.605EG + 0.882EGM. \quad (15)$$

Since the parameters {EG, EGM} in model (15) are inconvenient for data interpretation, the complicate parameter set {EG, EM, GM, EGM} can be legitimately replaced by the set {EG, EM, GM}, that is, the EGM term may be omitted with a negligible LR chi-square 0.089 ($df = 1$, $p = 0.765$). This leads to a simpler model with deviance 28.119 ($df = 23, p = 0.211$):

$$\text{Logit}[f(L|A, E, G, M, T)] = 0.790 - 0.733E - 0.388M - 0.471G + 0.487T - 0.352A$$
$$+ 0.815EM + 0.644EG + 0.142*GM. \quad (16)$$

Now, model (16) is simply reducible to the parsimonious AIC model (11) when the insignificant GM interaction ($p = 0.942$) is deleted. So far, the above analyses have shown that by inserting the insignificant EGM interaction ($\alpha_1 = 0.028$, $p = 0.049$) in Table 1 into the MI model (14), it would lead to locating the AIC model (11), which is near to the minimum AIC model (12).

In summary, the proposed two-step selection scheme outlined in Figure 1 is recommended for constructing parsimonious MI logit models as well as for finding the AIC models when the same five regressors (excluding C) are used. The constructed MI model (14) is shown to be more parsimonious than the AIC models (11) and (12) such that the factor G "gender" can offer unique main-effect data interpretation that is not given by the AIC models. In contrast, models (11) and (12) have extra significant interaction parameter estimates EG and {EG, MT}, respectively, hence smaller residual deviances than model (14); and therefore, they would be more frequently fitted to random subsets of the raw data, as compared with model (11).

A simulation study of 10,000 replicates of various subsample sizes of the raw data is conducted to examine the goodness-of-fit of the models (11), (12) and (14) under two types of sampling design. The first part assumes sampling under the true MI models (11), (12) and (14), respectively; and, the second assumes sampling random subsets of various sizes of the raw data without replacement. The simulation results are listed in Table 2 below.

Table 2. Proportions of accepting model (11), (12) or (14) under given sampling conditions

| Model/sample size \Assumption | True model (11) | True model (12) | True model (14) | Raw data random subsets |
|---|---|---|---|---|
| Model (11) / 800 | .9259 | .9112 | .9280 | .9266 |
| Model (12) / 800 | .9266 | **.9263** | .9273 | **.9395** |
| Model (14) / 800 | .8812 | .8676 | .9278 | .8921 |
| Mod (11) / 1000 | **.9249** | .9030 | .9302 | .9311 |
| Mod (12) / 1000 | .9248 | **.9243** | .9277 | **.9423** |
| Mod (14) / 1000 | .8737 | .8422 | **.9308** | .8896 |
| Mod (11) / 1500 | **.9275** | .8903 | .9288 | .9320 |
| Mod (12) / 1500 | .9255 | **.9307** | .9283 | **.9500** |
| Mod (14) / 1500 | .8416 | .7871 | **.9324** | .8669 |

It is seen in Table 2 that relatively higher proportions of test validity are acquired under each assumed true model against other models. It is expected that in accordance with the optimal prediction accuracy, the minimum AIC model (12) yields the highest proportions of test validity under the design of selecting random subsets of the raw data.

## 5. Concluding Remarks

The goal of this study is to present the MI approach to analyzing contingency table data, with particular emphasis on the logistic model selection. For the method of variable selection, the proposed selection scheme differs from the conventional regression analysis of the Type-III likelihood ratio statistics by evaluating the CMI statistics between each potential regressor and the target variable, conditional on the unselected regressors. In the proposed scheme for model selection, the selected regressors are rearranged to yield the least number of significant higher-order interaction effects, which is a new and useful idea in the literature. The main contribution of the proposed variable and model selection scheme lies in the design of deleting dispensable regressors and testing indispensable interaction effects, which lead to a straightforward construction of logistic models. The data example shows that the proposed MI analysis is able to conclude the unique EM interaction effect giving rise to the most parsimonious MI logistic model. Then, it also shows that the minimum AIC model can be easily identified in a neighbor of the EM model by using the same set of five regressors {A, T, G, M, E}. While binary variables are examined with logistic regression in the data analysis of the present study, generalization to multinomial variables is simply straightforward.

Recently, several studies suggested discretization of continuous variables for feature selection in regression analysis (Fan and Lo, 2013). The authors also indicated a tradeoff between information losses due to discretization versus information gains from robust detection of interactions by discretization. The empirical results demonstrated that the gain from robust detection of interactions suffices to offset the information loss due to discretization. On the other hand, studies using logistic regression also recommended rescaling interval variables into ordinal ones, for example, the family income is grouped into "below the poverty", "1-3.99 times the poverty income", and "4 or more times the poverty income" (Sullins, 2015). In principle, the mutual information identity introduced in Section 2 is valid for the analysis of categorical variables as well as continuous variables through discretization, that is, it is adaptable to the case of continuous regressors. We conclude the discussion with the recommendation of using the mutual information identities in the analysis of generalized linear models when both continuous and categorical variables are considered.

**Acknowledgment**

This study was supported by the grant MOST-105-2811-H-001-021 and by a special Visiting Research Project 103-2-1-06-14 while Professor C.C. Lin visited the Institute of Statistical Science, Academia Sinica in 2014.

**Background Documents**:

This paper was registered with the identification "arXiv:1801.01003" after it was rejected by *PLOS ONE* (November 23, 2017) without any comments on the authors' appeal (rebuttal, July 20, 2017) against a poor-level review on the initial manuscript (May 14, 2017). The first version of this paper was rejected by *Statistical Methods and Applications* (January 14 – April 27, 2015) using extraneous comparison with LASSO. It was also rejected by the editor of Methodology (Aug. 19, 2016) with his comments "… I find the issue of selection higher-order interactions in logistic regression not to be an urgent problem, …". But, he did not read our paper that "we showed: how to identify and delete insignificant high-order interactions".

**A Remark**:

All these reviews by SMA, Methodology and PLOS ONE ignored the fact that "the proposed information identity is by far the first method to directly construct the most parsimonious (logistic) regression model." For a bright scholar, he/she should realize that when a simple scheme of model selection is illustrated with a data example, it is sufficient to recognize what is the BEST method of model selection, whatever the dimension of the predictors. Thus, these editors and AEs appear to assert that "colleagues of their journals shall not use any ideas and/or methods of our inventions of information identity in this study, throughout the future: $21^{st}$ century".

Philip Cheng (retired Research Fellow, Academia Sinica)　　　　　April, 2018